# FreeSense:Indoor Human Identification with WiFi Signals

Tong Xin, Bin Guo, Zhu Wang, Mingyang Li, Zhiwen Yu
School of Computer Science
Northwestern Polytechnical University
Xi'an, P. R. China
guob@nwpu.edu.cn

*Abstract*—Human identification plays an important role in human-computer interaction. There have been numerous methods proposed for human identification (e.g., face recognition, gait recognition, fingerprint identification, etc.). While these methods could be very useful under different conditions, they also suffer from certain shortcomings (e.g., user privacy, sensing coverage range). In this paper, we propose a novel approach for human identification, which leverages WIFI signals to enable non-intrusive human identification in domestic environments. It is based on the observation that each person has specific influence patterns to the surrounding WIFI signal while moving indoors, regarding their body shape characteristics and motion patterns. The influence can be captured by the Channel State Information (CSI) time series of WIFI. Specifically, a combination of Principal Component Analysis (PCA), Discrete Wavelet Transform (DWT) and Dynamic Time Warping (DTW) techniques is used for CSI waveform-based human identification. We implemented the system in a 6m*5m smart home environment and recruited 9 users for data collection and evaluation. Experimental results indicate that the identification accuracy is about 88.9% to 94.5% when the candidate user set changes from 6 to 2, showing that the proposed human identification method is effective in domestic environments.

## I. INTRODUCTION

Generally, human identification is based on one or more intrinsic physiological [1-4] or behavioral [5-7] distinctions, which either is related to the shape of the body (e.g., fingerprint, iris, palm print, face characters) or certain behavior patterns of a person (e.g., typing, gait, voice rhythms). It plays a significant role in the area of pervasive computing and human-computer interaction. Currently, fingerprint- [2], iris- [4], and vein-based methods [8] have been successfully employed in automatic human identification systems. However, these systems require the user to be close to the sensing device for accurate identification. Researchers also make numerous attempts to develop methods for behavioral biometrics (mainly gait analysis) using cameras, radars, or wearable sensors [7]. However, the vision-based approaches only work with line-of-sight coverage and rich-lighting environments, which also cause privacy concerns. The low cost 60 GHz radar solutions can only offer an operation range of tens of centimeters [9], and the devices are not widely deployed in our daily life. Finally, wearable sensor-based approaches require people to wear some extra sensors.

WIFI techniques have been widely used in our daily life. Due to its popularity and low-deployment-cost, numerous researchers have devoted to pervasive sensing using WIFI devices, such as indoor localization [10], gesture recognition [11], etc. These studies are mainly based on RSSI, i.e., the coarse-grained signal strength information. Quite recently, Channel State Information (CSI), i.e., fine-grained information regarding WIFI communication become available. Specifically, CSI describes how the signal propagates from the transmitter to the receiver and reflects the combined effects of the surrounding objects (e.g., scattering, fading, and power decay with distance). There are many subcarriers in CSI, each of which contains the information of attenuation and phase shift. Therefore, CSI contains rich information and is more sensitive to environmental variances caused by moving objects [12]. Several notable studies on pervasive sensing have been conducted using CSI, such as high-accuracy human localization [13], human activity recognition [14-18], and crowd counting [12]. This paper, however, presents a new application based on CSI sensing — human identification.

Due to the difference of body shapes and motion patterns, each person can have specific influence patterns on surrounding WIFI signals while she moves indoors, generating a unique pattern on the CSI time series of the WIFI device. In this paper, we propose a novel approach called FreeSense, which can identify human indoors based on CSI-enriched WIFI devices. It is supposed to work in home environments (usually 2-6 family members) and deliver personalized services when recognizing the identity of a family member. There are two technical challenges faced by FreeSense, as presented below.

- Identification condition settlement and CSI time series segmentation. When a person moves around in the house, the influence level over multi-path communication of WIFI signals can be different when people walk across different paths. It is easy to understand that the identification performance can be raised at high human-WIFI-signal influence level, where the characterizing features are more distinguishable. In other words, it is difficult to identify human using the waveform caused by arbitrary walking of human. Therefore, we should find a proper functional window to maximize the identification performance. The waveform obtained within the functional window will be used for human identification.

- CSI feature extraction for human identification. Different people have different influence patterns to WIFI signals while moving around, regarding their body shapes and motion patterns. To conduct exact human identification over the extracted line-of-sight (LOS) waveform, we should utilize proper features to characterize the influences.

Ordinary features such as rate of change, signal energy, and maximum peak power are not effective, as they all have similar values for different people. The extraction of useful CSI features, however, becomes another challenge for this work.

To address these issues, FreeSense makes the following contributions:

- We propose a novel approach called FreeSense to identify human indoors based on WIFI CSI signals, which is non-intrusive and privacy-preserving compared with existing methods.
- We identify the line-of-sight path crossing moments as the functional window for enhanced human identification. Moreover, we put forward an algorithm to segment the CSI time series for the extraction of LOS waveforms within the functional window. It is based on the observation that the LOS waveform caused by human walking across LOS path shows a typical increasing and decreasing trend in rates of change in CSI time series.
- We use a model to extract the feature of LOS waveform, which consists of PCA and DWT. The recognition is based on the difference of personal movement influence to WIFI signals. We first use PCA to get the principal component of CSI time series, and then adopt DWT to compress the waveform and extract the shape feature of LOS waveform.

We deployed a prototype system in a 6m*5m smart home environment and recruited 9 users for evaluation. The results show that we can achieve 92.6% detection rate for CSI functional window segmentation, and achieve 88.9% to 94.5% accuracy when the candidate user set changes from 6 to 2, which is effective in domestic environments.

## II. RELATED WORK

### A. HUMAN IDENTIFICATION

Researchers have made numerous attempts to develop methods for human identification, which can be grouped into two categories: physiological feature based approaches and behavioral feature based approaches.

**Physiological features.** Fingerprint [2], iris [4], and vein authentications [8] have been successfully employed in automatic human identification systems. Furthermore, Duta [1] used the hand-shape to distinguish human. Chellappa et al. [19] proposed a method to recognize human based on face recognition. However, these systems require the subject to be close to the sensor for accurate identification.

**Behavioral features.** There are also some studies about behavioral biometrics, especially in gait analysis. Since human walking motion results in a self-sustaining and dynamic rhythm owing to the integrated signals generated from the spinal cord and sensory feedback, it contains unique characteristics of each individual. Little and Boyd [6] developed a model-free description of instantaneous motion and used it to recognize individuals by their gait. The work is based on camera sensing, which requires line of sight and enough lighting, and it also causes privacy issues. Nickel et al. [7] used HMM to recognize human gait based on accelerometer data, and it requires users to wear relevant sensors. We use WIFI signal to recognize human, which can provide better coverage and are device-free.

### B. CSI-BASED MOTION DETECTION

CSI values can be obtained from COTS WIFI network interface cards (NICs) (such as Intel 5300 [20] and Atheros 9390 [21]). It is used for human activity recognition and indoor localization. Han et al. proposed WiFall, which can detect falls of human in indoor environments by analyzing CSI values [15]. E-eyes proposed by Wang et al. exploits CSI for recognizing household activities such as washing dishes and taking a shower [18]. Nandakumar et al. leveraged the CSI and RSS information offered by off-the-shelf WIFI devices to classify four arm gestures - push, pull, lever, and punch [16]. Wang et al. proposed WiHear, which use CSI to recognize the shape of mouth while human speaking [17]. Ali et al. proposed WiKey, which can recognize keystrokes by using CSI [14]. These studies are all about CSI-based human activity recognition. Our work, however, concentrates on human identification, which is not investigated in existing CSI-based studies.

## III. PROBLEM ANALYSIS AND SYSTEM FRAMEWORK

### A. PROBLEM ANALYSIS

#### 1) Identification condition settlement

To enhance the identification performance, it is important to determine the functional window in CSI sensing. Based on observations from a large amount of data, we find that the WIFI signal has a more notable fluctuation when people walk across the LOS path. Figure 1 shows the amplitudes of CSI time series of five different subcarriers of a transmit-receive antenna pair, where a person repeatedly walking across the LOS path. Compared with Non-Line-Of-Sight (NLOS) paths, the signal of LOS path is more stable [22]. Thus, we propose to identify human by analyzing waveform caused when people walking across the LOS path. Specifically, we call this waveform line-of-sight (LOS) waveform.

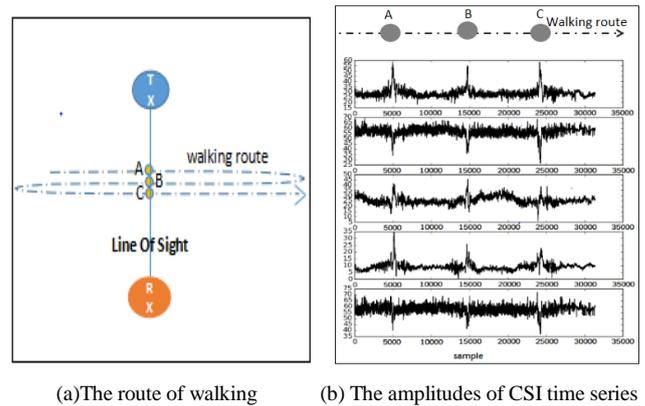

(a) The route of walking     (b) The amplitudes of CSI time series

Fig. 1. The waveform of walking

#### 2) CSI time series segmentation

After defining the functional window, i.e., the line-of-sight waveform, we need to extract it from the whole CSI time series. We find that the waveforms caused by human walking across LOS path have a typical increasing and decreasing trend in rates of change in CSI time series. Based on this characteristic, WiKey [14] proposed an algorithm based on mean absolute deviation (MAD) to segment the CSI time series. However, it will cause high error rate due to the interference waveform caused by human walking across other paths, and cannot be directly used in our work. Instead, while we also design the segmentation algorithm based on this characteristic, a sliding window approach is adopted to search the point, which has an increasing and decreasing trend before or after it. Moreover, to reduce the error rate, we add more qualifications. The detail of our algorithm will be given in Section 4.

*3) Feature Extraction*

According to experimental observations, we find that the unique human influence on WIFI signal may reflect on both time and frequency domain features. Therefore, we use shapes of the extracted line-of-sight waveform as features, which retains both time and frequency domain information of the waveform.

Each transmit-receive (*TX-RX*) antenna pair of a transmitter and receiver has 30 subcarriers. Let $M_{Tx}$ and $M_{Rx}$ represent the number of transmitting and receiving antennas. There are 30 * $M_{Tx}$ * $M_{Rx}$ CSI streams in a time-series of CSI values. Directly using the 30 * $M_{Tx}$ * $M_{Rx}$ CSI streams will lead to high computational cost. We thus need to compress the data. As Figure 1 shows, all subcarriers show correlated variations in their time series. Thus, we use Principal Component Analysis (PCA) to reduce the dimensionality of the CSI time series. Subsequently, we also apply Discrete Wavelet Transform (DWT) to compress the extracted line-of-sight waveform, which can preserve most of the time and frequency domain information.

*4) Classification Method*

The starting time and ending time of Line-of-sight waveforms determined by the extraction algorithm are not exact, which make the midpoints of extracted line-of-sight waveforms rarely align with each other. Meanwhile, the lengths of different waveforms caused by different human walking are also different. Thus, two waveforms cannot be compared using standard measures like correlation coefficient or Euclidean distance. To address this issue, we adopt the Dynamic Time Warping (DTW) technique, which can find the minimum distance alignment between two waveforms of different lengths.

### B. SYSTEM FRAMEWORK

As shown in Figure 2, our system consists of three parts: Data Collection and Noise Removing, Feature Extraction, and Human Identification.

**Data Collection and Noise Removing.** We first record the CSI value with COTS WIFI network interface cards (NICS), while people walk across the LOS path. Afterwards, we choose a suitable filter to remove noise.

**Feature Extraction.** We extract shape features of line-of-sight waveforms from the CSI time series, which can be used to recognize human identity.

**Human Identification.** Based on the extracted features, a k-nearest neighbor (KNN) classifier is used for human identification.

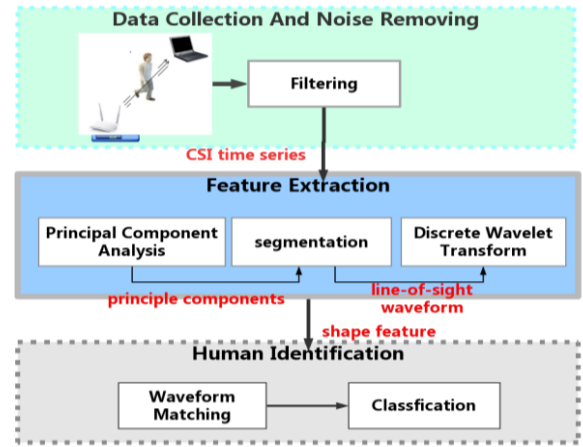

Fig. 2. The FreeSense Framework

### IV. DETAILED DESIGN OF THE SYSTEM

#### A. NOISE REMOVING

The CSI values provided by commodity WIFI NICS are inherently noisy. Thereby, we should remove the noise at first. Specifically, the Butterworth IIR filter is adopted, which does not significantly distort the phase information in the signal and has a maximally flat amplitude response in the passband. According to experiments, we find that the frequency *f* of the variations in CSI time series caused by human walking is around 10Hz. As the sample rate of CSI data is 1000Hz, the cut-off frequency $w_c$ of Butterworth filter is calculated as formula (1):

$$w_c = \frac{2\pi * f}{F_s} \approx 0.06 rad/s \qquad (1)$$

The filtering result is shown in Figure 3(b).

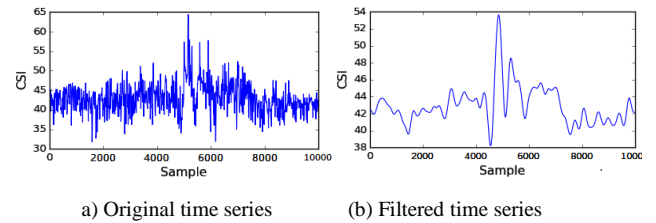

a) Original time series      (b) Filtered time series

Fig. 3. Original and filtered CSI time series.

#### B. FEATURE EXTRACTION

To extract feature from the CSI time series, we perform PCA in each *TX-RX* antenna pair to reduce the dimension at first. Then, we segment the CSI time series to identify the starting time and ending time of each line-of-sight waveform. Finally, we adopt DWT to compress these waveforms.

*1) Principal Component Analysis*

For each *TX-RX* antenna pair, we perform PCA to get *p* principle components. As a result, we have $p * M_{Tx} * M_{Rx}$

waveform shapes for each human. Based on the observation that only the top 4 principle components contain significant variations in CSI values caused by human walking, we set $p=4$. Figure 4 shows the top 4 principle components of one TX-RX antenna pair. PCA can lead to different ordering of principal components in waveforms of different walking of the same person, because the ordering of waveforms done by PCA is based solely on the value of their variance. In order to minimize the possibility of reordering, we order the projected waveforms in a descending order of their peak to peak values before using the waveforms for feature extraction and classifier training.

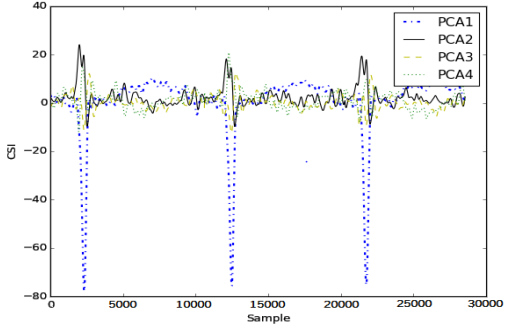

Fig. 4. PCA of CSI stream

*2) CSI time series segmentation*

To obtain the line-of-sight waveform from CSI time series, we need identify the start time and end time of the waveform. Specifically, based on the observation that the line-of-sight waveform of human walking across LOS path shows a typical increasing and decreasing trend in rates of change in CSI time series, we put forward a segmentation algorithm consisting the following four steps:

First, for every $j$-th point of principle components $1 \leq k \leq p$, the algorithm calculates the mean absolute deviation (MAD) of each window of size $w$ before and after it as formula (2).

$$MAD_{before}^{j}[k] = \frac{\sum_{i=j}^{j+w} |Y(i) - \overline{Y}(j-w:j)|}{w}$$
$$MAD_{after}^{j}[k] = \frac{\sum_{i=j}^{j+w} |Y(i) - \overline{Y}(j:j+w)|}{w} \quad (2)$$

Second, we calculate the sum of MAD of all $p$ principle components in $j$-th point as formula (3).

$$MAD_{before}^{j} = \sum_{k=1}^{p} MAD_{before}^{j}[k]$$
$$MAD_{after}^{j} = \sum_{k=1}^{p} MAD_{after}^{j}[k] \quad (3)$$

Third, the algorithm set thresholds $T_1$ and $T_2$, $T_1 > T_2$. For every start point $j_{begin}$ and end point $j_{end}$, it should satisfy the condition as formulated in (4):

$$MAD_{before}^{j_{begin}} \leq T_2, \quad MAD_{after}^{j_{begin}} \geq T_1$$
$$MAD_{after}^{j_{end}} \leq T_2, \quad MAD_{before}^{j_{end}} \geq T_1 \quad (4)$$

Finally, we have got some start point and end point already. But not all point is correct. Thus, we introduce the following limitations:

1. We record the end point only if we have got a start point;

2. $timelen_1 \leq j_{end} - j_{begin} \leq timelen_2$ (5)

where $timelen_1$ and $timelen_2$ are two thresholds, which are empirically determined. So far, we get the point pair $< j_{begin}, j_{end} >$, based on which we can obtain the Line-of-sight waveform.

*3) Discrete Wavelet Transform*

We use DWT to compress every line-of-sight waveform, which can preserve most of the time and frequency domain information. Specifically, according to experimental results, we choose the Daubechies D4 wavelet and use approximation coefficients to represent the waveform shape features.

C. HUMAN IDENTIFICATION

We design a $k$-nearest neighbor (KNN) classifier based on DTW. Specifically, given two line-of-sight waveform series, $X$ and $Y$, the distance between them is calculated as formula (6):

$$dis(X,Y) = \sum_{m=1}^{M_T * M_R} \sum_{k=1}^{p} DTW_m(X_k, Y_k) , \quad (6)$$

where $p$ is the dimension of the line-of-sight waveform of each TX-RX antenna pair, and $M_T * M_R$ is the number of antenna pairs. The classifier searches for the majority class labels among $k$ nearest neighbors based on the corresponding shape features using the DTW distance metric.

V. EVALUATION AND DISCUSSIONS

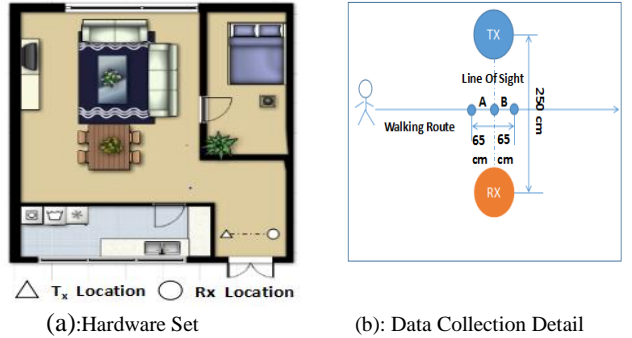

(a): Hardware Set  (b): Data Collection Detail

Fig. 5. Experimental Scene: Hardware Set

A. IMPLEMENTATION

*1) Hardware Setup*

We used a Lenovo X200 laptop with Intel Link 5300 WIFI NIC as the receiver, which has 4GB of RAM and Ubuntu 14.04 as its operating system. TP-Link TLWR1043ND WIFI router was adopted as the transmitter operating in 802.11n AP mode at 2.4GHz. Specifically, the transmitter has 2 antennas and the receiver has 3 antennas, i.e., $M_T = 2$ and $M_R = 3$.

As shown in Figure 5(a), we implemented our experiment in a typical home environment. We placed the receiver and transmitter on the entry hall with a distance of 250cm. The CSI values are measured on ICMP ping packets sent from the WIFI router to the laptop at the data rate of about 1000 packets/s.

*2) Data Collection*

We recruited 9 people as volunteers to implement our experiments, who are all university students at the age of 20 to 26. Each of the volunteers had provided 40 samples.

To increase comparability of the result, we set some limits for their walking steps, as shown in Figure 5(b). We asked volunteers walk across the LOS path from left to right and adjust their starting point to ensure that they walked in a normal status, and put down the left foot in area A and the right foot in area B. Certainly, we can identify human for more walking conditions, if we extend the train sets.

### B. LINE-OF-SIGHT WAVEFORM DETECTION ACCURACY

TABLE I. ACCURACY FOR SEGMENTATION

| Method | WiKey | Our method |
|---|---|---|
| Detection Ratio | 91.7% | 92.6% |

TABLE II. ERROR RATE FOR SEGMENTATION

| Method | WiKey | Our method |
|---|---|---|
| Error Ratio | 16.2% | 6.5% |

We adopt the WiKey algorithm as the baseline, and compared it with our algorithm based on two metrics: Detection Ratio (*DR*) and Error Ratio (*ER*), which are defined as formula (7) :

$$DR = \frac{N_{cd}}{M_a} \times 100\%, \quad ER = \frac{N_{fd}}{M_d} \times 100\% \quad (7)$$

where $N_{cd}$ represents the number of correctly detected line-of-sight waveforms, $M_a$ is the total number of all line-of-sight waveforms caused by human walking, $N_{fd}$ is the number of falsely detected line-of-sight waveforms, and $M_d$ stands for the total number of all detected line-of-sight waveforms. We find that both of the two algorithms obtained high detection ratios, while our algorithm had a lower error ratio.

### C. CLASSIFICATION ACCURACY

*1) Accuracy for different number of human*

We used 20 samples for each person to train the classifier, and the rest samples for testing. The identification accuracy achieved 75.5% to 94.5% while the number of users varied from 2 to 9. Specifically, the accuracy will decline with the increase of users, as shown in Figure 6. The reason is that as more people who have similar body shapes or motion patterns are involved in the system, the difficulty of identification would increased. It is worth to mention that as most of the family sizes range from 2 to 6, the identification accuracy (up to 88.9%) is effective and acceptable for family use in home environments.

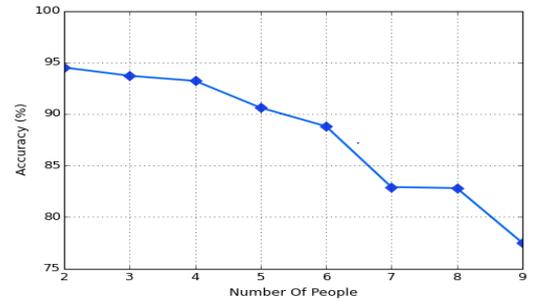

Fig. 6. Accuracy for 2-9 people

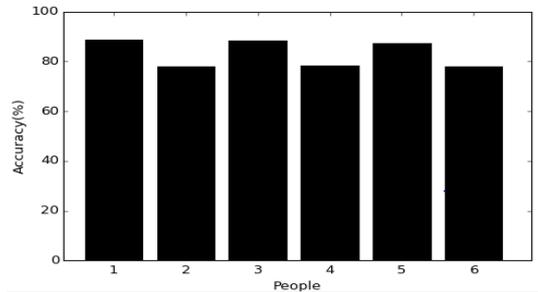

Fig. 7. Performance for a round trip

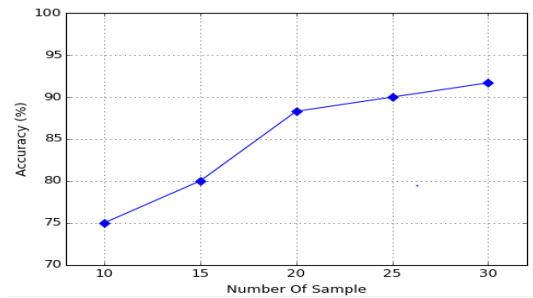

Fig. 8. Effect of the size of training set

*2) Performance for more walking conditions*

We further present the performance of the system for more walking conditions. We let the volunteers walk a round trip along the walking route, asking them to put down the left foot in area B at first and then put down the right one in area A when they walked across LOS path from right to left. Similarly, we also asked them to put down left foot in area A at first and then put down the right one on area B when they walked across LOS path from left to right. Figure 7 shows the accuracy of 6 people in this condition. The system can achieve nearly 81.5% accuracy on average.

*3) Effect of the size of training sets*

To determine the impact of the number of training samples on the accuracy, we test the accuracy with different size of training sets.

According to Figure 8, the accuracy increased when the number of training samples changed from 10 to 30, where the corresponding recognition accuracy increased from 75.0% to 91.7%.

*D. DISCUSSIONS*

We have reported our preliminary efforts on indoor human identification with WIFI signals in this work. In this section, we discuss several limitations and opportunities to improve our work in the future.

According to the experimental results, we can achieve more than 88% accuracy with 6 subjects as family members. Though, the accuracy might be affected by several factors, such as the distance between the subject and the transmitter, crossing manners of the LOS path, and so on. As an early effort, we make some restrictions in the experiments. In the future, we plan to make more experiments to study the detailed impacts of such factors and improve our approach.

Surrounding people may also have influence on the accuracy of FreeSense. Currently, we mainly test our work under one-person environments. In the future, we will test its performance when there are more than one person in the same space, study its effective bounds, and improve the human identification performance.

## VI. CONCLUSION

In this work, we put forward a novel approach called FreeSense to identify human indoors based on WIFI CSI signals. Compared with existing studies, it has merits such as non-intrusive and privacy-preserving. We propose a model to extract the feature of line-of-sight waveforms, which contain the information of human walking. To evaluate the system, we design and implement a set of experiments. The results show that the detection rate is about 92.6%, and the identification accuracy ranges from 94.5% to 88.9% when the number of users changes from 2 to 6, which is effective in domestic environments. We intend to extend our work by measuring the impact of different variables and improving the identification performance.

## VII. ACKNOWLEDGMENTS

This work was partially supported by the National Basic Research Program of China (No.2015CB352400), the National Natural Science Foundation of China (No. 61332005, 61373119, 61402369).